\documentclass[journal]{IEEEtran}

\usepackage{amsmath,amsfonts}
\usepackage{algorithmic}
\usepackage{algorithm}
\usepackage{textcomp}
\usepackage{stfloats}
\usepackage{url}
\usepackage{verbatim}
\usepackage{graphicx}
\usepackage{cite}
\usepackage{color}
\usepackage[tight]{subfigure}

\usepackage{acronym}
\newacro{gfm}[GFM]{Grid-Forming}
\newacro{gfl}[GFL]{Grid-Following}
\newacro{ibr}[IBR]{Inverter-Based Resource}
\newacro{pll}[PLL]{Phase-Locked Loop}
\newacro{vsm}[VSM]{Virtual Synchronous Machine}
\newacro{sm}[SM]{Synchronous Machine}
\newacro{cf}[CF]{Complex Frequency}
\newacro{coi}[CoI]{Center of Inertia}
\newacro{rocof}[RoCoF]{Rate of Change of Frequency}
\newacro{dfig}[DFIG]{Doubly-Fed Induction Generator}
\newacro{gvr}[GVR]{Global Voltage Regulation}
\newacro{qss}[QSS]{Quasi-Steady State}
\newacro{pmu}[PMU]{Phasor Measurement Unit}

\begin{document}

\title{Frequency Quality Metrics based on Second-Order Derivative and Autocorrelation}

\author{Taulant K\"{e}r{\c{c}}i,~\IEEEmembership{Senior Member,~IEEE}, and Federico Milano,~\IEEEmembership{Fellow,~IEEE}
   \thanks{T.~K{\"e}r{\c c}i is with the Irish Transmission System Operator,
     EirGrid, Ballsbridge, D04FW28, Ireland.}
   \thanks{F.~Milano is with School of Electrical and Electronic Engineering,
    University College Dublin, Belfield Campus, D04V1W8, Ireland.
    Corresponding author's e-mail: federico.milano@ucd.ie.}
  
  \thanks{This work was partially supported by Sustainable Energy Authority of Ireland (SEAI) by funding F.~Milano through FRESLIPS project, Grant No.~RDD/00681.}
  \vspace{-5mm}
}

\maketitle

\begin{abstract}
   
This industry-oriented paper originates from the observation that current frequency quality metrics utilized by transmission system operators (TSOs) fail to fully capture the dynamic behavior of the grid frequency.  Motivated by this gap, the paper proposes novel frequency quality metrics based on second-order dynamics and stochastic autocorrelation.  Using real-world data with 0.1 s and 1 s resolution from the Irish, Great Britain and Nordic systems and running dynamic stochastic simulations, the paper shows that the proposed metrics bring new and counterintuitive insights in terms of how good or poor the frequency quality of power grids is beyond current well-known metrics.  In particular, the paper shows that a power system may show good frequency quality using standard metrics and poor frequency quality using the proposed metrics.  
\end{abstract}


\section{Introduction}
\label{sec:intro}

Frequency quality is a topic of huge interest to transmission system operators (TSOs) \cite{kercci2026frequency}.  To quantify it,  TSOs utilize several different metrics.  Among them, minutes outside defined ranges such as $\pm$100 mHz or $\pm$200 mHz, standard deviation of the frequency, $\sigma_{f}$, and rate of change of frequency (RoCoF) are few of the most important ones.   While these metrics are extremely useful for TSOs, they do not capture important dynamics of system frequency such as its second-order dynamics and autocorrelation.  The latter describe important features of stochastic processes such as temporal dependencies and memory.  

In terms of power system applications, references \cite{9019633, 8963682, 9637935, stochastic, schafer2018non} are among the very few works that utilize the second-order derivative of frequency or its autocorrelation function (ACF).  In particular, while the authors in \cite{8963682, stochastic, schafer2018non} analyze the ACF of some real-world grids, including of the All-Island power system (AIPS) of Ireland, Great Britain (GB) and Nordic, they do not discuss what additional information ACF brings compared to existing metrics aimed at quantifying frequency quality.  In this context, this industry-oriented work brings the following novel contributions:
\begin{itemize}
    \item Propose novel frequency quality metrics based on second-order derivatives of system frequency and its ACF.  These metrics are specifically aimed at quantifying the impact of  fast dynamics on frequency quality.
    \item Show through real-world data and dynamic stochastic simulations that the proposed metrics provide useful operational insights to TSOs and complement the information obtained with conventional metrics.
\end{itemize}

\section{Conventional Metrics}
\label{sec:old_metrics}

Table~\ref{tab:qualitymetrics} presents conventional frequency quality metrics and their operational limits, if any, for the AIPS, GB and the Nordic grids.  These conventional metrics, or variants of them, are commonly utilized to evaluate the impact of different resources and/or control strategies on frequency quality \cite{MALAMAKI2023101211}.

The standard deviation ($\sigma_{f}$) of a time series is defined as:
\begin{equation}
  \label{eq:sigma1}
  \sigma_{f} = \sqrt{ \frac{1}{N} \sum \limits_{t=1}^{N} (f_t - \overline{f})^2 } \, , \\
\end{equation}
where $f_t$ represents the instantaneous frequency at time $t$, $\overline{f}$ represents the mean frequency, and $N$ represents the total number of samples of the frequency time series.  

RoCoF is calculated as follows:
\begin{equation}
  \label{eq:rocof}
  \mathrm{RoCoF} = \frac{f_{t}-f_{t-\tau}}{\tau} \, ,
\end{equation}
where $\tau$ represents the time interval. 

\begin{table}[htb]
  \centering
  \caption{Key frequency quality metrics utilized by TSOs} 
  \label{tab:qualitymetrics}
  \resizebox{0.8\linewidth}{!}{
  \begin{tabular}{lcccccc}
    \hline
     System & Minutes outside $\pm$100, $\pm$200 mHz & $\sigma_{f}$ & \rm RoCoF \\
     & (mins) & (Hz) & (Hz/s) \\
     \hline
    AIPS & 15,000 ($\pm$200 mHz) & -- & 1 \\
    GB & 15,000 ($\pm$200 mHz) & -- & 0.5 \\
    Nordic & 15,000 ($\pm$100 mHz) & -- & --\\
    \hline
  \end{tabular}}
\end{table}

\vspace{-3mm}

\section{Proposed Metrics}
\label{sec:new_metrics}

This section defines the two metrics that constitute the main contribution of this work.  

\subsubsection{Rate of Change of RoCoF}

As indicated above, RoCoF is an important and widely used metric for TSOs.  However, due to the rapid growth of highly variable demand and generation technologies, it appears relevant to also calculate its derivative in order to capture fast and nonlinear dynamics.  For this reason, we define $\mathrm{RoCoF'}$ as follows:
\begin{equation}
  \label{eq:rocof_rocof}
  \mathrm{RoCoF'} = \frac{\rm RoCoF_{t}^{\tau}-\rm RoCoF_{t-\Delta\tau}^{\tau}}{\Delta\tau} \, ,
\end{equation}
where, similar to $\tau$, $\Delta\tau$ is a time window parameter for $\mathrm{RoCoF'}$ calculation.  To quantify and get an understanding of the long-term stochastic evolution of second-order derivatives, we calculate the standard deviation of $\mathrm{RoCoF'}$, $\sigma_{\mathrm{RoCoF'}}$:
\begin{equation}
  \label{eq:sigma2}
  \sigma_{\mathrm{RoCoF'}} = \sqrt{ \frac{1}{N-2} \sum \limits_{t=1}^{N-2} (\mathrm{RoCoF'}_t - \mathrm{\overline{RoCoF}'})^2 } \, . 
\end{equation}

\subsubsection{Autocorrelation of system frequency}

The ACF of a stochastic process, in this case, of the frequency time series data, is the measure of correlation of the current values to the past values of the frequency.  It measures the linear dependence of the frequency to the delayed version of the same frequency over progressive time delays.  This aspect of frequency quality is not captured with existing metrics and, as shown below, contributes to an increased understanding of frequency quality.  This is consistent with the findings of recent literature, such as \cite{9637935}, which shows that a high value of the autocorrelation of a stochastic process can lead to high variations of the system variables and, in some cases, to instability.  

The ACF can be expressed as a function of time delay $\theta$, and is written as follows:
\begin{equation}
\label{eq:auto}
R_{\kappa_f}(\theta) = \frac{\mathrm{E}[(\kappa_f(t) - \mu_{\kappa_f})(\kappa_f(t + \theta) - \mu_{\kappa_f})]}{\sigma_{\kappa_f}^{2}} \, ,
\end{equation}
where $R_{\kappa_f}$ is the ACF of the stochastic process $\kappa_f$; and $\mu_{\kappa_f}$ and $\sigma_{\kappa_f}^{2}$ are the mean and variance of $\kappa_f$, respectively.

In order to quantify frequency quality based on $R_{\kappa_f}(\theta)$ we need to fit it to relevant functions such to a sum of damped sinusoidal and decaying exponential functions and then use the parameters of these functions such as the exponential parameter $\alpha$, as important frequency quality metrics.  In this work, the fitted ACF ($R_{\kappa_f}(\theta)$) is approximated as:
\begin{equation}
\label{eq:fit}
    R_{\kappa_f}^{\rm fitted}(\theta) = \underbrace{u_1 e^{-\alpha_{\text{fast}}\theta}}_{\text{term}_1} + \underbrace{(1 - u_1) e^{-\alpha_{\text{slow}}\theta} \cos(\omega \theta)}_{\text{term}_2} \, ,
\end{equation}
where $u_1$ represents the weight of the initial fast decay component; $\alpha_{\text{fast}}$ represents the decay rate of the initial ``shock'' drop; $\alpha_{\text{slow}}$ represents the decay rate of the oscillatory component; and $\omega$ represents the angular frequency of the oscillation.  Equation~\eqref{eq:fit} is fitted to $R_{\kappa_f}(\theta)$ by utilizing a non-linear least squares method, included in the Python package SciPy \cite{JONSDOTTIR2019368}.  

As illustrated in the next sections, \eqref{eq:fit} fits well the ACF of power system frequency.  From a frequency quality perspective lower values of $\alpha_{\text{fast}}$, $\alpha_{\text{slow}}$, and $\omega$ indicate good frequency quality as it means frequency time series values are correlated and do not change fast in different timescales ($\alpha_{\text{fast}}$, $\alpha_{\text{slow}}$), as well as do not oscillate ($\omega$).  The limit case is a steady-state/constant frequency which leads to zero values for all parameters.  Conversely, high values of $\alpha_{\text{fast}}$, $\alpha_{\text{slow}}$, and $\omega$ indicate poor frequency quality.

\section{Real-World Data}
\label{sec:real}

To calculate the proposed frequency quality metrics, we utilize AIPS, GB, and Nordic frequency time-series data from September 2025 and, for comparison, some data from December 2025, with 1 s resolution.   It has been shown in the literature that one month data and analysis is sufficient to generalize frequency quality behavior \cite{11299077}.  As available data have 1 s resolution, we assume static time intervals to calculate RoCoF and $\mathrm{RoCoF'}$, that is, $\tau$ = 1 s and $\Delta\tau$ = 1 s in \eqref{eq:rocof} and \eqref{eq:rocof_rocof}, respectively.  For comparison, we also utilize 0.5 s static time intervals for the calculation of the different RoCoFs for September 2025 given the availability of 0.1 s data resolution from the AIPS and Nordic grid but not from the GB.  It is worth stressing that while industry practice is to calculate RoCoF over a 0.5 s rolling window, the European Union regulation recognizes 1 s resolution as valid for frequency quality evaluation purposes, as follows \cite{sogl}: ``\textit{‘instantaneous frequency data’ means a set of data measurements of the overall system frequency for the synchronous area with a measurement period equal to or shorter than one second used for system frequency quality evaluation purposes;}''.

\color{black}
\vspace{-3mm}

\subsection{September 2025}
\label{sec:september}

\subsubsection{1 s resolution and 1 s sampling time\color{black}}
\label{sec:september1s}

\begin{figure}[thb!]
  \begin{center}
    \resizebox{0.8\linewidth}{!}{\includegraphics{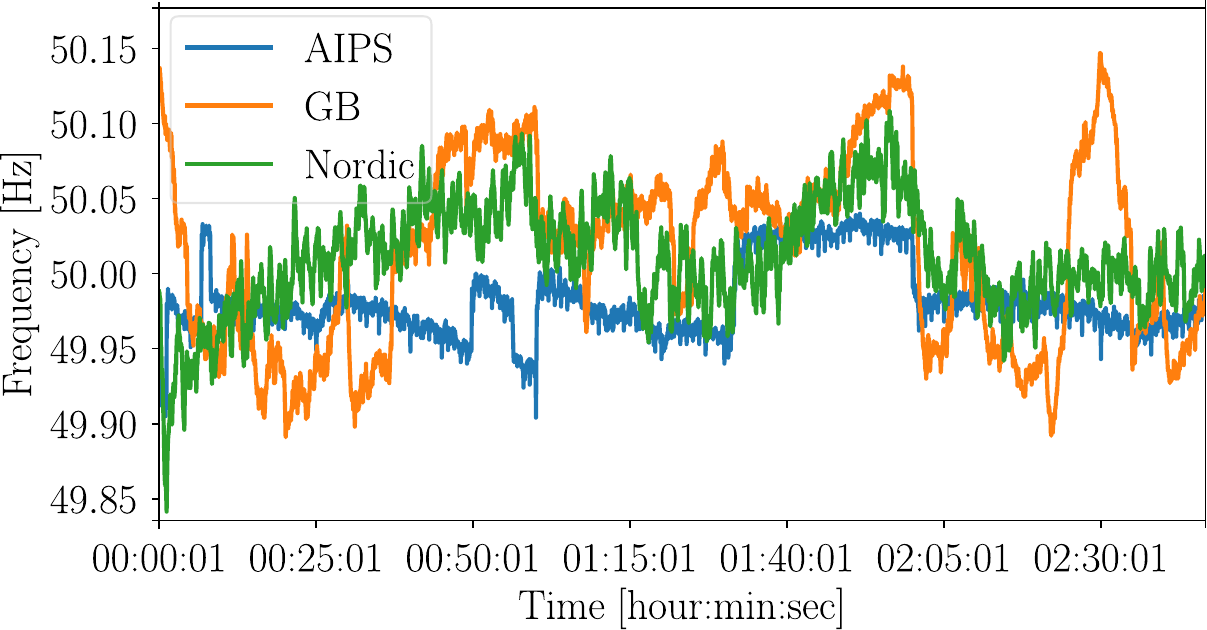}}
    \caption{Frequency variations for a selected random time window.}
    \label{fig:freq}
  \end{center}
  \vspace{-4mm}
\end{figure}

Figure~\ref{fig:freq} depicts relevant frequency traces of the three power systems using, for illustration purposes, the first 10,000 data points of 1st September 2025 (random window selection).  All three systems show stochastic behavior of system frequency with relevant jumps as well.  In particular, the GB grid, despite being much bigger than AIPS and slightly lower than Nordic, demonstrates more volatile frequency around 50 Hz.

Table~\ref{tab:auto_september} shows that the Nordic grid exhibits superior frequency quality when referring to conventional metrics ($\sigma_{f}$ = 0.032 Hz).  Similar conclusion may be reached if we calculate the minutes outside defined ranges (see Table~\ref{tab:qualitymetrics}) where the AIPS, GB and the Nordic demonstrate 0, 8.91, and 1.55 minutes, respectively.  This is due to the Nordic TSOs implementing automatic generation control (AGC), whereas AIPS and GB implement a manual secondary frequency regulation.  However, if we calculate the standard deviation of the ${\rm RoCoF}$ ($\sigma_{\mathrm{RoCoF}}$) and $\sigma_{\mathrm{RoCoF'}}$ the situation changes. That is, the performance of Nordic is low with respect to the proposed metric ($\sigma_{\mathrm{RoCoF'}}$ = 0.0040 $\rm Hz/s^{2}$).  This is a counterintuitive result given the Nordic grid has a bigger size and lower $\sigma_{f}$ than AIPS and GB. Based on a time series of 10,000 data points, the Nordic grid shows the highest $\sigma_{\mathrm{RoCoF'}}$ (0.0036 $\rm Hz/s^{2}$) compared to the AIPS ($\sigma_{\mathrm{RoCoF'}}$ = 0.0017 $\rm Hz/s^{2}$) and GB ($\sigma_{\mathrm{RoCoF'}}$ = 0.0027 $\rm Hz/s^{2}$).  Note that due to its relatively slow timescales, the AGC is effective in reducing $\sigma_f$ but does not damp and/or reduce the negative effect of oscillations on frequency quality ($\sigma_{\mathrm{RoCoF'}}$).

\begin{table}[h!]
  \centering
  \caption{Frequency quality metrics for September 2025} 
  \label{tab:auto_september}
  \resizebox{0.9\linewidth}{!}{
  \begin{tabular}{lcccccccccc}
    \hline
     System & $\sigma_f$ & $\sigma_{\mathrm{RoCoF}}$ & $\sigma_{\mathrm{RoCoF'}}$ & $u_1$ & $\alpha_{\text{fast}}$ & $\alpha_{\text{slow}}$ & $\omega$   \\
     & (Hz) & (Hz/s) &($\rm Hz/s^{2}$) \\
     \hline
    AIPS & 0.041 & 0.0022 & 0.0020 & 0.3931 & 0.0003 & 0.0013 & 0.0012  \\
    GB & 0.074 & 0.0024 & 0.0029  & 0.2249 & 0.0003 & 0.0017 & 0.0013  \\
    Nordic & 0.032 & 0.0031 & 0.0040 & 0.7016 & 0.0016 & 0.0422 & 0.0419  \\
    \hline
  \end{tabular}}
\end{table}

\begin{figure}[htb]
  \centering
  \subfigure[September 2025]{\includegraphics[width=0.775\linewidth]{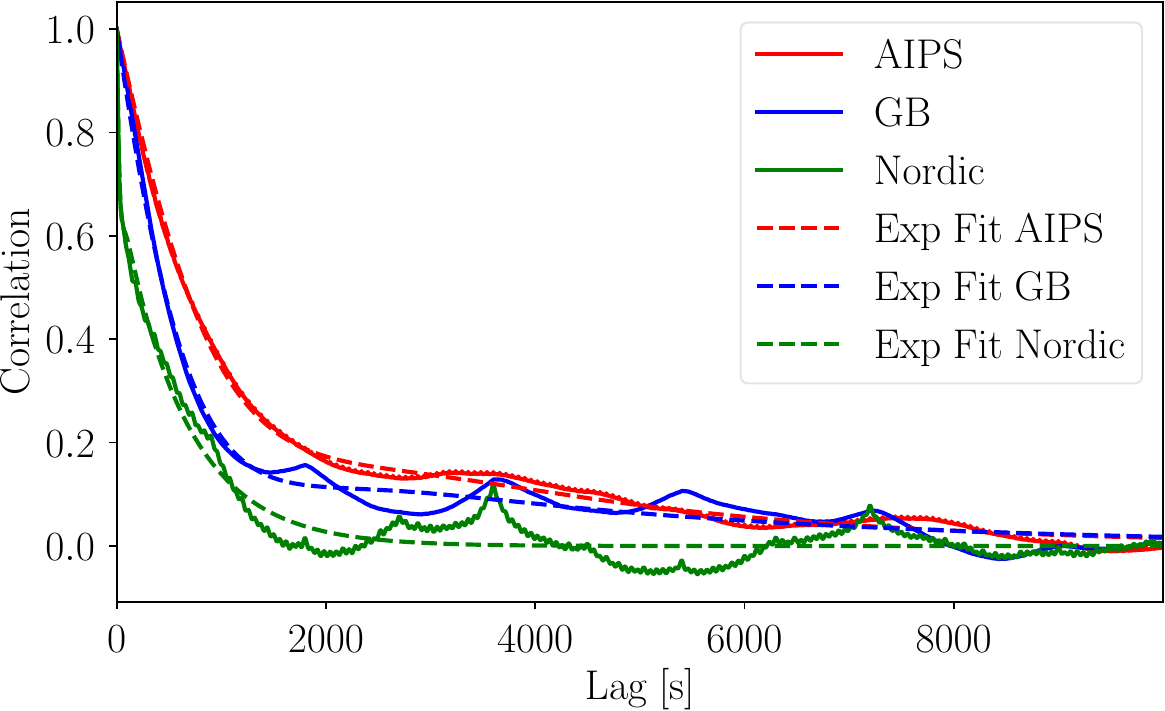}} \\
  \subfigure[December 2025]{\includegraphics[width=0.775\linewidth]{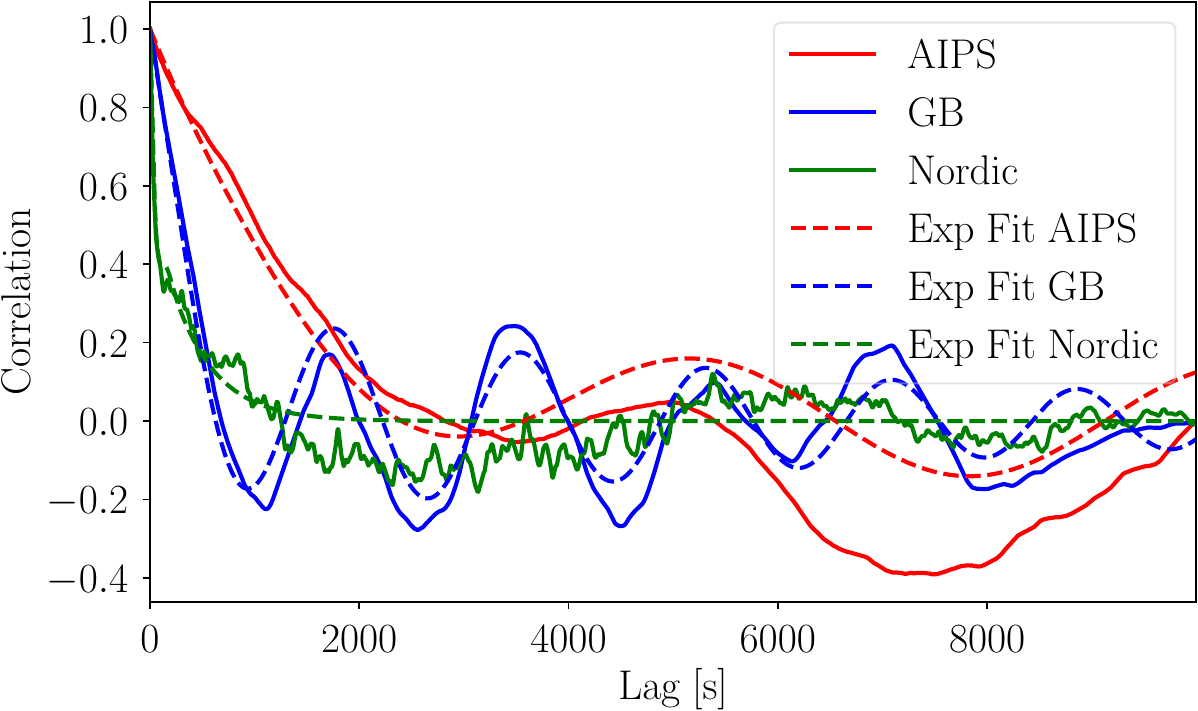}}
  \caption{ACFs of the frequency and their fitted versions for real-world data.}
  \label{fig:autocorr_september}
\end{figure}

Figure \ref{fig:autocorr_september} supports this conclusion by showing that the Nordic grid has a very fast decaying ACF in the initial time lags.  Relevant parameters of the fitting function \eqref{eq:fit} in Table~\ref{tab:auto_september} namely $\alpha_{\text{fast}}$ and $\alpha_{\text{slow}}$ confirm these results by showing higher values compared to the GB and AIPS.  This result is consistent as, despite all three power systems have significant share of variable wind and solar generation, the AIPS has a lower level of load volatility relative to its size compared to what is the case in GB and the Nordics (e.g., have more embedded generation and heavy industry).  As a matter of fact, it has been shown in the literature that the rate of change of demand is one of the main causes of frequency events in GB \cite{HOMAN2021116723}.  Similarly, load fluctuations with a wide frequency range (i.e., approximated to white noise) are one of the main sources of ultra low-frequency oscillations in the Nordic grid \cite{entsoe}.  

Figure \ref{fig:autocorr_september} also reveals correlation peaks every 15/30 minutes for GB and the Nordic synchronous areas which indicate periodicity in the data.  These peaks are due to market-driven imbalances that are more pronounced in self-dispatch systems (Nordic and GB) rather than central-dispatch ones (AIPS).  

\subsubsection{0.1 s resolution and 0.5 s sampling time\color{black}}
\label{sec:september01s}

This section discusses the effect of data resolution and shorter time intervals on the proposed metrics.  With this aim, all metrics are calculated again using a 0.1 s sampling resolution from the AIPS and Nordic (no similar data from the GB was available) and 0.5 s time interval to calculate RoCoF-based metrics.  

\begin{table}[b!]
  \centering
  \caption{Frequency quality metrics for September 2025 using 0.1 s resolution and 0.5 s time interval for RoCoF metrics.} 
  \label{tab:auto_september_100}
  \resizebox{0.9\linewidth}{!}{
  \begin{tabular}{lcccccccccc}
    \hline
     System & $\sigma_f$ & $\sigma_{\mathrm{RoCoF}}$ & $\sigma_{\mathrm{RoCoF'}}$ & $u_1$ & $\alpha_{\text{fast}}$ & $\alpha_{\text{slow}}$ & $\omega$   \\
     & (Hz) & (Hz/s) &($\rm Hz/s^{2}$) \\
     \hline
    AIPS & 0.041 & 0.0032 & 0.0086 & 0.40 & 0.0000 & 0.0001 & 0.0001  \\
    Nordic & 0.034 & 0.0028 & 0.097 & 0.64 & 0.0002 & 0.0063 & 0.0043  \\
    \hline
  \end{tabular}}
\end{table}

Results are shown in Table~\ref{tab:auto_september_100}.  Except for a small difference in $\sigma_{\mathrm{RoCoF}}$, these results are consistent with those of Table~\ref{tab:auto_september}.  That is, the Nordic grid shows lower $\sigma_{f}$ but higher $\sigma_{\mathrm{RoCoF'}}$ and higher $\alpha_{\text{fast}}$, $\alpha_{\text{slow}}$, and $\omega$ compared to the AIPS.  While using different time series resolution and/or different sampling times, general conclusions do not change.

\color{black}

\vspace{-3mm}

\subsection{December 2025}
\label{sec:december}

To complement the analysis of the previous section, we select another random period namely the first 10,000 data points from 1st December 2025 and calcualte the proposed metrics.  Figure~\ref{fig:autocorr_september} and Table~\ref{tab:auto_december} confirm the previous results and show that the Nordic grid exhibits a rapid decaying ACF and higher values of $\alpha_{\text{fast}}$ and $\alpha_{\text{slow}}$ compared to the AIPS and GB power systems.  This happens despite the Nordic grid shows lower $\sigma_{f}$ (0.023 Hz) compared to the other two systems.  All three systems demonstrate 0 minutes outside defined ranges.

\begin{table}[htb]
  \centering
  \caption{Frequency quality metrics for December 2025} 
  \label{tab:auto_december}
  \resizebox{0.9\linewidth}{!}{
  \begin{tabular}{lcccccccccc}
    \hline
     System & $\sigma_f$ & $\sigma_{\mathrm{RoCoF}}$ & $\sigma_{\mathrm{RoCoF'}}$ & $u_1$ & $\alpha_{\text{fast}}$ & $\alpha_{\text{slow}}$ & $\omega$   \\
     & (Hz) & (Hz/s) &($\rm Hz/s^{2}$) \\
     \hline
    AIPS & 0.041 & 0.0015 & 0.0014 & 0.857 & 0.0008 & 0.0000 & 0.0012  \\
    GB & 0.059 & 0.0027 & 0.0041  & 0.7100 & 0.0024 & 0.0001 & 0.0035  \\
    Nordic & 0.023 & 0.0018 & 0.0019 & 0.5798 & 0.0025 & 0.0207 & 0.0297  \\
    \hline
  \end{tabular}}
\end{table}

Compared to Table~\ref{tab:auto_september}, Table~\ref{tab:auto_december} shows that $\sigma_{\mathrm{RoCoF}}$ and $\sigma_{\mathrm{RoCoF'}}$ for Nordic are lower (0.0018 and 0.0019 $\rm Hz/s^{2}$, respectively) than $\sigma_{\mathrm{RoCoF}}$ and $\sigma_{\mathrm{RoCoF'}}$ for GB (0.0027 and 0.0041 $\rm Hz/s^{2}$, respectively).  This indicates higher rate of change of demand in GB.   Moreover, the GB system shows a higher oscillating behavior indicating strong periodicity in the time series data.  A way to address this poor frequency quality can be, for the GB TSO, to increase the volumes of procured reserves.  The proposed metrics, thus, complement each other and provide additional information to TSOs beyond existing conventional metrics.

\section{Case Study}
\label{sec:case}

\begin{figure}[t!]
  \begin{center}
    \resizebox{0.8\linewidth}{!}{\includegraphics{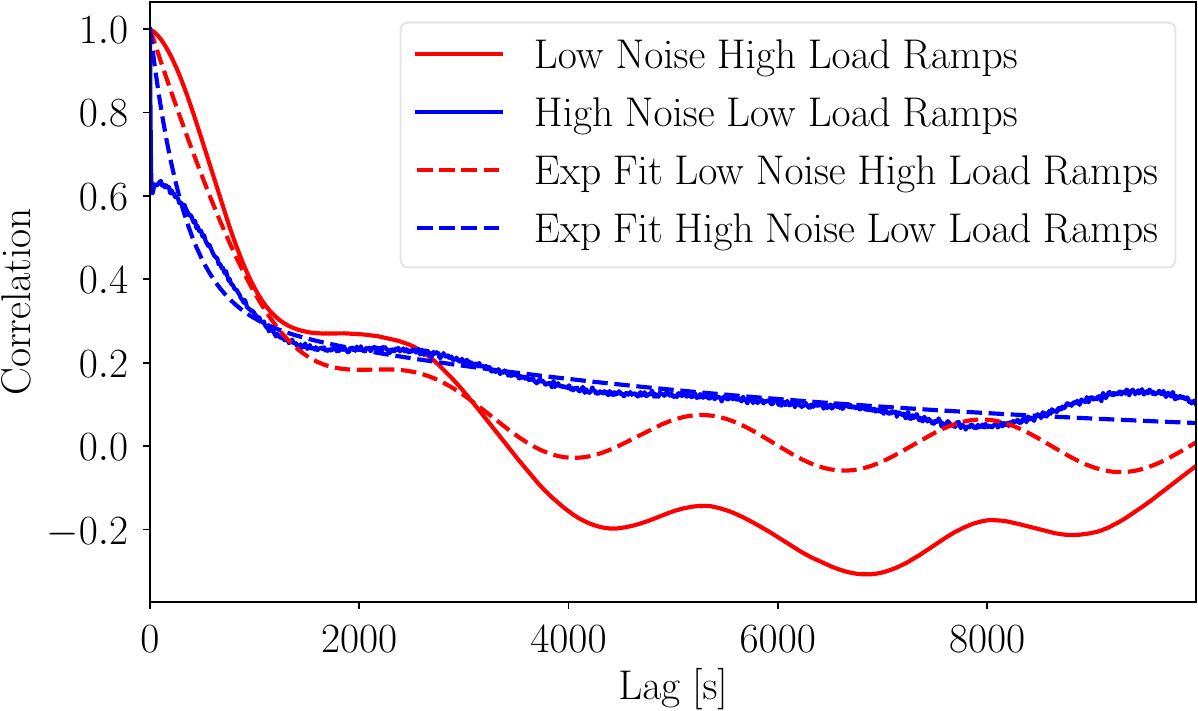}}
    \caption{ACFs of the frequency for simulations and their fitted versions.}
    \label{fig:case}
  \end{center}
  \vspace{-4mm}
\end{figure}

To validate the conclusions drawn based on real-world data, we employ the IEEE 9-bus system and run 24 h dynamic stochastic simulations using Dome \cite{6672387}.  Two scenarios are simulated namely one with high noise and relatively low load ramps (e.g., Nordic) and one with low noise and high load ramps (e.g., AIPS).  Primary frequency control is the same in both cases and no AGC is assumed for simplicity.  Figure~\ref{fig:case} and Table~\ref{tab:case} confirm the real-world results (e.g., Section \ref{sec:december}).  That is, while the case with low noise and high load ramps leads to higher $\sigma_{f}$ (0.032 Hz) it also shows lower $\sigma_{\mathrm{RoCoF'}}$ (0.0010 $\rm Hz/s^{2}$) and $\alpha_{\text{slow}}$ (0.0000).  Conversely, the case with high noise and low load ramps demonstrates lower $\sigma_{f}$ (0.015 Hz) but higher $\sigma_{\mathrm{RoCoF'}}$ (0.0041 $\rm Hz/s^{2}$) and, in turn, shows a rapid decaying ACF (first 10,000 data points) and higher $\alpha_{\text{slow}}$ (0.0030).  Finally, both scenarios show 0 minutes outside $\pm$200 mHz range.

\begin{table}[t!]
  \centering
  \caption{Frequency quality metrics for the stochastic simulations} 
  \label{tab:case}
  \resizebox{1.0\linewidth}{!}{
  \begin{tabular}{llcccccccccc}
    \hline
     Noise & Ramps & $\sigma_f$ & $\sigma_{\mathrm{RoCoF}}$ & $\sigma_{\mathrm{RoCoF'}}$ & $u_1$ & $\alpha_{\text{fast}}$ & $\alpha_{\text{slow}}$ & $\omega$   \\
     & & (Hz) & (Hz/s) &($\rm Hz/s^{2}$) \\
     \hline
    Low & High & 0.032 & 0.00093 & 0.0010 & 0.9374 & 0.0008 & 0.0000 & 0.0024  \\
    High & Low & 0.015 & 0.0038 & 0.0041  & 0.3277 & 0.0002 & 0.0030 & 0.0000  \\
    \hline
  \end{tabular}}
  \vspace{-2mm}
\end{table}

\section{Conclusions}
\label{sec:conclu}


The letter proposes novel metrics based on the second-order dynamics and autocorrelation of the frequency and evaluates these new metrics based on operational data from the AIPS, GB and Nordic, as well as to dynamic stochastic simulations.  It is shown that the proposed metrics contribute to an increased understanding of frequency quality.  For example, the letter shows that a power system can demonstrate good frequency quality w.r.t. the existing metrics and poor frequency quality w.r.t. the proposed metrics (e.g., low $\sigma_{f}$ but high $\sigma_{\mathrm{RoCoF'}}$).  

Future work will focus on evaluating and record over a reporting period the proposed metrics in real-world power systems with the goal of identifying the sources of poor frequency quality.  We will also develop practical countermeasures to improve frequency quality such as novel controllers and reserve services.

\bibliographystyle{IEEEtran}
\bibliography{refs}

\end{document}